\documentclass[twocolumn,showpacs,preprintnumbers,amsmath,amssymb]{revtex4}
\usepackage{graphicx,psfrag,bbm,latexsym,color,dcolumn,bm,dsfont,bbm,color,mathrsfs,bbold,latexsym,amsmath,amsfonts,amssymb}

\newcommand{\beq}{\begin{equation}}
\newcommand{\eeq}{\end{equation}}

\newcommand{\beqa}{\begin{eqnarray}}
\newcommand{\eeqa}{\end{eqnarray}}

\def \as {\relax\ifmmode\alpha_s\else{$\alpha_s${ }}\fi}

\begin{document}

\title{Exact Nonperturbative Renormalization}
\author{P.R. Crompton}
\affiliation{Institut f\"ur Theoretische Physik I., Universit\"at Hamburg, Jungiustrasse 9, D-20355 Hamburg, Germany}
\affiliation{Center for Theoretical Physics, Massachusetts Institute of Technology,
77 Massachusetts Ave., Building NE25 4-111, Cambridge, MA  02139, USA.}
\vspace{0.2in}
\date{\today}

\begin{abstract}
\vspace{0.2in}
{We propose an exact renormalization group equation for Lattice Gauge Theories, that has no dependence on the lattice spacing. We instead relate the lattice spacing properties directly to the continuum convergence of the support of each local plaquette. Equivalently, this is formulated as a convergence prescription for a characteristic polynomial in the gauge coupling that allows the exact meromorphic continuation of a nonperturbative system arbitrarily close to the continuum limit. }
\vspace{0.1in}
\end{abstract}


\maketitle

\section{introduction}

Many physical problems in gauge field theories involve performing integrations over multiple scales, and in particular, relating these differing scales between perturbative and nonperturbative regimes. This is the picture
of asymptotic freedom \cite{Gross}. For example, we may have a system of vector bosons with some large invariant mass, $Q^{2}$, at small transverse momentum, $q^2$, as is the case with Drell-Yan processes \cite{soft}. In the region $Q^2 \gg q^2 \gg \Lambda_{\rm QCD}^2$ perturbative methods will yield convergent solutions, but for small $q^2$ nonperturbative correction effects will become important \cite{resum}. Typically, we would want to investigate the relative perturbative and nonperturbative scale dependences for various quantities of physical interest, $F(Q^{2},g)$, by evaluating the renormalization group equation of a system as a function of, $g$, the local loop vertex interaction strength. This definition of renormalization, and its determination between scales is a central element in extracting data from numerical simulations for all manner of experimentally relevant functions, $F(Q^{2},g)$, such as meson decay constants, form factors, structure functions, and mixing amplitudes \cite{npr}.

The simplest expression of this renormalization group equation, valid for both perturbative and nonperturbative scales, is given by \cite{Gell},

\beq
\label{Gell}
\frac{\partial F(Q^{2}/\mu^{2},g_{\mu}) }{\partial \mu}
\equiv \left( x\frac{\partial }{\partial x} - \beta\frac{\partial }{\partial 
g}\right) F(x,g)
= 0\,,
\eeq
where, $x\equiv Q^{2}/\mu^{2}$, and, $\mu$, is some arbitrary reference point or physical scale \cite{strong}. The generic solution of this renormalization group equation will yield a characteristic equation for the function,
$F(Q^{2},g)$. The group theoretic interpretation of this property is as a mapping relation for the scale transformations, where the basis of the solutions of (\ref{Gell}) form the generators of the group of the scale transformation symmetries. Implicit to this relation is also $\beta$, a local coupling expansion term \cite{beta} for the bases. Simplistically, difficulties arise in relating perturbative and nonperturbative regimes when these local coupling expansions in $g$ are not in comparable bases for the two regimes. The respective regimes in which loop propagator expansions yield well-defined analytic solution cannot then be simply mapped into each other, and the same renormalization prescription cannot then be applied to both problems. The local effective coupling constants of one regime will, in general, contain nonlocal IR divergences compared to the local couplings of the loop operator expansion in the other regime \cite{dr1}-\cite{dr3}. 

Typically, relating perturbative and nonperturbative scales involves defining a nonperturbative renormalization prescription from a numerical Lattice Gauge Theory input. For an arbitrary physical observable calculable as a matrix element as a function of the local gauge field operators, $\{U\}$, this renormalization is defined through,
\beq
F(a,\mu) = c_W(\mu) \,Z(a \mu) \,\, \langle U \vert O(a) \vert U' \rangle\,,
\eeq

where $c_W$ is the Wilson coefficient \cite{sf1}\cite{npr2}, $a$ is the lattice spacing, $Z$ the renormalization constant responsible for relating the perturbative and nonperturbative scales, and $\mu$ is the momentum dependence of the quark states the operators interpolate between. The relationship between $c_W$ and $Z$ is hopefully as comparable local expansions, or equivalently mutually comparable bases for the characteristic polynomial defined in (\ref{Gell}). The assumptions for believing this can be satisfied, and such a region can be found in $\mu$ and $a$ essentially follows from the definition of the lattice renormalization group equation, and the local effective 
expansion for the usual lattice gauge fields definition \cite{strong}\cite{lattice}\cite{sym}. What we seek to do here, which is different from previous nonperturbative renormalization schemes in which this region for (2) is investigated, is to treat the effective local expansion of Lattice Gauge Theory as a nonlocal, noncommutative 
problem that cannot be factored into a convergent expansion in the lattice spacing. We presently ignore improvement to $\mathcal{O}(a^{2})$ although the new prescription we define can be equally well generalised for such renormalization group improved lattice definitions.

Our new scheme will proceed by defining the dual of a numerically realised lattice system, on a generic level, and then relating the continuum convergence of local coupling expansions for both the physical lattice system and its dual to each other. This will lead to an exact nonperturbative renormalization group equation, of the form of (1), as a finite characteristic polynomial. We then give a simple, practical gauge-fixing prescription to identify the nonlocal basis of this expansion from the operator elements of a conventional Lattice Gauge Theory simulation. The advantage of the approach, and the reason for developing a new noncommutative formalism for Lattice Gauge Theory, is that in this new basis the lattice system will be meromorphically continuable in $g$ towards the continuum limit of 
vanishing lattice spacing. In this new formalism the lattice $\beta$-function will be implicitly absorbed into the local effective expansion and, in principle, this can be used to completely remove the main source of systematic uncertainties in relating perturbative and nonperturbative scales through the lattice spacing.

The article is organized as follows. In Section II we review existing nonperturbative renormalization methods
in Lattice Gauge Theory, following from Symanzik's improvement program, and specifically focus on their lattice spacing dependence. In Section III we give a definition for our noncommutative lattice gauge operators, and relate these directly to the operators of standard lattice actions. We also give an overview of the conditions for the generic convergence of the zeroes of a characteristic polynomial formed over the lattice sites in an expansion of these operators. This defines the basic convergence properties of the characteristic equation that we will use to describe our new lattice renormalization group equation. In Section IV we explicitly evaluate the converse Mellin transform that defines our lattice partition function dual in this formalism. This defines the form that Wilson propagators must take on the the lattice in order to be meromorphically continuable in the lattice cutoff, and consequently, the UV regularization of the approach. In Section V we compare this UV regularized characteristic polynomial form directly with the series based $\zeta$-function renormalization procedure for noncommutative quantum field theories. In Section VI, we solve our new renormalization group equations to show that all divergences in the Wilson propagators can be completely expressed as the zeroes of the characteristic polynomial of the physical lattice, following a nonlocal gauge fixing prescription for the noncommutative operator definitions. In Section VII we finally apply the new procedure to logarithmically divergent and power divergent operators in the renormalization of four-fermion lattice interactions for $\Delta S=2$ and $\Delta S=1/2$ processes, and comment on how the new 
prescription compares with existing approaches in the continuum limit.

\section{Nonperturbative Renormalization}
In this article we will consider an unimproved definition of lattice gauge fields, following Wilson:

\beq
\label{Wilson}
U(n;\nu) = \exp[i az A_\nu(n)] \,\, ;  \quad A_\nu(n) = \sum_{d=1}^{8} 
A_\nu^d(n)\lambda^d,
\eeq
where $U$ is the lattice gauge field, $n$ the lattice site index, $z$ the gauge coupling, $A$ the continuum
gauge field, $\nu$ the spatial orientation of the lattice gauge field, and the $\{\lambda\}$ the usual Gell-Mann
matrices for $SU(3)$ \cite{lattice}. Lorentz invariance for these standard Lattice Gauge Theory elements is only recovered in the limit of vanishing lattice spacing, and strictly the elements form a noncommuting series expansion in the generators of the SU(3) algebra with the expansion term being the lattice spacing.

What Symanzik suggested, in a series of articles, was that local effective expansion was possible for Lattice Gauge Theory, and could be used as the basis for a renormalization group equation in the lattice spacing \cite{sym}. What was done, schematically, was to write down a generalised effective action for a continuum theory, add a linear correction, and then argue that both actions separately have a valid continuum renormalization prescription, by induction. The new insight we have, follows from our result in \cite{parity}. Here we 
have identified nonlocal solutions to the groundstate of the generalised effective action. These nonlocal solutions arise, not from the physical poles, but the branch points associated with resolving global symmetries on a local level. What we have identified is that the minima of the free energy of a partition function may not necessarily correspond to the saddle-point solution of a system, which is the assumption behind Symanzik's argument.

The definition of a generic operator renormalization constant $Z$ from nonperturbative studies is also essentially derived from this continuum picture of Symanzik. The fact that the two continuum theories in this picture share the same renormalization prescription is used to justify continuing the Lattice Gauge Theory elements in (3) in the lattice spacing. The analyticity of $Z$ in (2) follows from this assumption. Three basic methods can be used to calculate $Z$; Lattice Perturbation Theory, Lattice Ward identities \cite{wi} or Schr\"odinger Functionals \cite{sf1}\cite{sf2}, and the Regularization Independent Momentum scheme (RIMOM), which consists in defining renormalization conditions on quark correlation functions \cite{rimom}. The basic underlying idea of these last two schemes is to define an improved $\mathcal{O}(a^{2})$ effective expansion from either improved operator definition, as Symanzik had suggested, BRST gauge fixing prescriptions, or a combination of both. This is used to find the dependence of $Z$ on $a$. This gauge fixing prescription follows from Becchi, Rouet, Stora, and (independently) Tyutin in \cite{brst}. What is not directly addressed in any of these existing renormalization prescriptions is how to completely remove this lattice spacing dependence and directly impose Lorentz invariance exactly. For this, we have argued in \cite{parity}, we need to consider a full noncommuting expansion. Whilst in the continuum Symanzik's picture for relating perturbed theories is valid, we have argued the convergence of the support of a finite lattice system to the continuum limit is not necessarily so well-defined, and we would need to give a modified lattice $\beta$-function definition that resolves some of the ambiguity associated with nonlocal symmetries to treat the problem of topological corrections to the IR \cite{dr1}-\cite{dr3}.

\section{Noncommuting Quantum Loop Operators}
In this article we will consider a quantum loop operator definition of the partition function $\mathcal{Z}$ defined over an extended phase space, which gives a noncommutative picture of the dynamics of the Wilson loop operators on the lattice, 
\beqa
\label{partitionz}
\mathcal{Z}_{N}(t) & = & \int {\rm{d}}g \,\,\,
{\rm{exp}}\!\left[\,\int_{0}^{t} B_g(\bm{n}_s)\,z-V(\bm{n}_s) \,\,ds 
\,\right] \\
& \equiv & \int {\rm{d}}g \,\,\, 
{\rm{exp}}\!\left[\,\int_{0}^{t} L(\bm{n}_s,z)\,z \,\,ds \,\right]\,,
\eeqa
where,

\beqa
\label{Hilberta}
B_g(\bm{n}) & \equiv &
\sum_{(i,j)}^{N \otimes T} \,\, \sum_{\sigma \in G}
\lambda_{ij\sigma}(\bm{n})\frac
{\langle \bm{n}\oplus \bm{1}_{i\sigma} \oplus \bm{1}_{j\sigma}| g \rangle}
{\langle \bm{n}| g \rangle}\,, \\
\label{Hilbertb}
V(\bm{n})  & \equiv &
\sum_{(i,j)}^{N \otimes T} \,\, \sum_{\sigma \in G}
\lambda'_{ij\sigma}(\bm{n})\frac
{\langle \bm{n}\oplus \bm{1}_{i\sigma} \oplus \bm{1}_{\sigma j}| \bm{n} 
\rangle}
{\langle \bm{n}| \bm{n} \rangle}\,,
\eeqa
with $t\in \mathbb{R}, \,\,\, z \in \mathbb{C}$, $\lambda_{ij\sigma}(\bm{n}),\lambda'_{ij\sigma}(\bm{n})\in \mathbb{C}$, $\{\sigma\}$ form some finite subset of the elements of $G$, $L$ is the loop operator defined in a form 
we will use later, and $g$ is a general element of the general continuous group $G$ \cite{presilla-gauge}\cite{presilla-old}. The meaning of this extended phase space $N\otimes T$ is quite simplistic. The usual definition for lattice gauge fields involves exponentiation, this is a noncommutative operation, and strictly speaking a local expansion can only be formed to leading order in the lattice spacing without further improvement of the operators or gauge-fixing. Here we very simply separate the lattice elements into commuting and noncommuting parts,

\beqa
U(s\,;s+1) & = & iazA_{\mu}(n) - (az)^{2}A^{2}_{\mu}(n) \,\,\, ...  
\nonumber \\
& = & z \, B_g(\bm{n}_s)\,-V(\bm{n}_s)\,.
\eeqa
The action in (4), with this choice in (8), is therefore of the form of a pure gauge theory, but with this formalism is more directly related to the continuum gauge fields. In particular, we can consider how $B$ evolves into $A$ by considering how the partition function will evolve as a holomorphic function of the gauge coupling $z$. The lattice spacing itself is also absent in this new notation. The reason for this is that we will implicitly absorb the lattice spacing into the operator definitions throughout the article, such that all the singularities of the lattice system will completely specified through the nonanalyticities of the gauge coupling $z$ and lattice cutoff $t$
including the lattice $\beta$-function. If we do want to consider any fermionic interaction terms, for a more general Lattice Gauge Theory action in (8) and (4), these interactions will be expressed uniquely through the symmetries of the second term, (7). However, even if there are no fermionic interaction terms the second operator will still represent exactly, in this formulation, all of the disconnected contributions to the partition function.

The extended dimension $T$ in this picture, has the meaning of the algebra of the space that is realised in practice on a finite lattice system. It is the Hilbert space of the noncommutative geometry associated with the disconnected contributions to the finite lattice system partition function. The purpose of rewriting the standard lattice gauge fields in this formalism is not envisaged as a basis for new numerical simulation techniques, but rather, to give a formulation for Wilson loop operators that will allow the lattice $\beta$-function to be reformulated as a finite expansion in the gauge coupling $z$ and lattice cutoff $t$.

From the loop operator definition in (\ref{Hilberta}) and (\ref{Hilbertb}) we can also equivalently define the finite volume partition function over the subspace $T$, with $z$ as the lattice cutoff, at least on a symbolic level,

\beqa
\label{partitiont}
\mathcal{Z}_{T}(z) & = & \int {\rm{d}}g \,\,\,
{\rm{exp}}\!\left[\,\int_{0}^{z} A_{g}(\bm{n}_{s'})\,t-V(\bm{n}_{s'}) 
\,ds'\,\right] \\
& \equiv & \int {\rm{d}}g \,\,\,
{\rm{exp}}\!\left[\,\int_{0}^{z} L(\bm{n}_{s'},t)\,t \,\,ds' \,\right]\,.
\eeqa
This partition function in (10) defines the dual of the Lattice Gauge Theory that is realised in practice via Markov chain dynamics.   In \cite{parity} we discussed the existence of the continuum limit for the two partition functions, 
which we have now defined in (\ref{partitionz}) and (\ref{partitiont}). The discussion in \cite{parity} was given in the context of the partition function zeroes of quantum systems. The defining property of quantum partition function zeroes, which are zeros of the characteristic polynomial formed over the lattice sites as an expansion in either $z$ or $t$ respectively for $\mathcal{Z}_{N}(t)$ and $\mathcal{Z}_{T}(z)$ is that aside from isolated singularities the partition function is holomorphic in the expansion variable. What we would like to do in this article is relate the convergence of these two partition function zeroes expansions.

The main conceptual difference in the approach we are proposing, from existing nonperturbative renormalization group 
methods, is that we do not now seek in this article to try to improve our knowledge of the local expansion in either the gauge coupling $z$ or the lattice spacing, as is being done in \cite{sf1}\cite{rimom}. Rather, we will try to identify the relationship between the two local expansions for the saddle-point equations in the physical and dual lattice systems, in order to exactly constrain the expansion in $z$ through the convergence of the expansion in $t$. The approach we will define is very similar to an asymptotic expansion about a renormalization group fixed point, but we will use the asymptotic convergence of two such expansions to constrain each other, and then define the renormalization group flow from nonperturbative input. In doing so we define an exact Renormalization Group equation by defining the characteristic polynomial for the lattice partition function defined over $N$ and its continuum convergence as a function of the gauge coupling $z$ by absorbing the lattice $\beta$-function into the operator definition in (6) and (7). 

\section{UV Regularization via a local Mellin transform of the plaquettes}

In \cite{parity} we deduced the asymptotic saddle-point solution of the partition function, $\mathcal{Z}_{N}(t)$, of a generic Lattice Gauge Theory by evaluating the inverse Laplace transform of the partition function in (4) expressed as a function of the poles of the meromorphic system, of the form, 
\beq
\label{LtransZ}
\mathcal{Z}_{N}(t) = \int_{0}^{\infty} e^{tz} \tilde{\mathcal{Z}}_{N}(z)\,\, 
  dz, \quad
\tilde{\mathcal{Z}}_{N}(z) = \prod_{k=0}^{N} \left(\frac{1} {z+V_{k}}
\right)\,. 
\eeq
If we make a Wick rotation in $z$ then this Laplace transform has a usual expression in terms of the poles of the general $N$-loop propagator of the system, defined in momentum space. In fact this gives a definition of the usual Wilson loop propagators in our formalism. However, unlike conventional momentum space definitions we have implicitly included an abelian phase, so that at the corners of the lattice Brillouin zone there will be no ambiguity associated with noncompact angles when the lattice system is UV regularized. Our loop operators in this picture do not simply appear as vortices and antivortices, but rather can have essentially unrestricted multiplicities in the continuum theory, and correspond to any topological sector. If we compare this Laplace transform definition for Wilson loops with a conventional Fourier transform definition in momentum space, on a finite system, we will be passing through a finite number of noncompact angles as we take the usual UV limit of the upper boundary condition in (\ref{LtransZ}). 

However, to define the analytic continuation that relates schemes through the Wick rotation we must simply only redefine the contours of the Laplace transforms. This contour redefinition must be that which encloses all the singularities arising from the branch points coming from passing through the compact phase an arbitrary number of times. We gave the general prescription in \cite{parity} for dealing with branch points in this Laplace transform formulation, which itself is a simple modification of the Residue theorem. 

Given that we can simply relate notations via contour redefinition we want to now explicitly evaluate the partition function of the dual $\mathcal{Z}_{T}(z)$ in the form of a product of poles. This is given by,

\beq
\label{LtransT}
\mathcal{Z}_{T}(z) = \int_{0}^{\infty} e^{zt} \tilde{\mathcal{Z}}_{T}(t)\,\, 
  dt, \quad
\tilde{\mathcal{Z}}_{T}(t) = \prod_{k'=0}^{T} \left(\frac{1} {t+V_{k'}} 
\right)\,.
\eeq
To be clear, (\ref{LtransT}) is not the Laplace transform of the partition function in (\ref{LtransZ}). The relationship between $\mathcal{Z}_{N}(t)$ and $\mathcal{Z}_{T}(z)$ is very complicated, and what we are trying to establish. Rather than a Laplace transform, it is more convenient to express (\ref{LtransT}) in terms of a Mellin transform, via the substitution $x=e^{-z}$. Whilst a Laplace transform definition involves giving a suitable contour $\mathcal{C}$ to avoid the singularities associated with branch cuts and poles, the analogue for the Mellin transform is the fundamental strip. This fundamental strip completely specifies the domain of analyticity of the Mellin transform, and is given by $-\alpha < {\rm{Re}}\,t < \beta$. The partition function of the dual in its form as a product of poles in (\ref{LtransT}) can therefore rewritten as,
\beq
\label{Mellin}
{\tilde{\mathcal{Z}}}_{T}(t) = \int_{0}^{\infty} x^{t-1} \, 
\mathcal{Z}_{T}(x)  \, dx\,,
\eeq
\beqa
\label{fs1}
\alpha & = & {\rm{sup}} \{ a : \mathcal{Z}_{T}(x) = \mathcal{O}(x^{a}), 
x\rightarrow 0^{+}\}\,, \\
\label{fs2}
\beta & = & {\rm{sup}} \{ b : \mathcal{Z}_{T}(x) = \mathcal{O}(x^{-b}), 
x\rightarrow \infty \} \,.
\eeqa
What we can immediately identify from this is that a finite order polynomial in $z$ defined over the subspace $N$ in (11), will constrain the domain of analyticity in $t$ in (12), and vice versa. So, if we try and write down a characteristic polynomial for the lattice in the local gauge coupling $z$ or the lattice spacing in order to define an exact nonperturbative renormalization group equation on the finite lattice system it will not be fully analytic. This characteristic polynomial will only be defined analytically in $z$ subject to the constraints coming from $t$. Whilst this is a general feature of lattice calculations and their effective scales, for this new noncommutative lattice formulation in which the noncompactness is resolved, we can directly relate the cutoff $t$ for the action to the lattice gauge coupling $z$ exactly for each individual local term, not just for the system as a whole as in (2), since the lattice $\beta$-function is absorbed into the definition of (11) and (12).

The converse mapping theorem for Mellin transforms can be used to formally evaluate the definition in (\ref{LtransT}) 
for arbitrary values of the cutoff of the dual $z$ to investigate the UV regularization of the physical lattice system. That is, defining the Wilson loops in terms of the dual and Mellin transforms, we can use the definitions in (\ref{fs1}) and (\ref{fs2}) to try and define the general polynomial form of the exact expansion in gauge coupling of one partition function, in terms of the other. The essence of the converse mapping theorem result is to show that the Mellin transform in (\ref{Mellin}) will admit a meromorphic continuation to some extended strip $\langle\gamma, \beta\rangle$ with $\gamma < \alpha$. The formal conditions for this to be fulfilled are that the extended strip contain a finite number of poles, be analytic on the boundary of the region at ${\rm{Re}}\,t = \gamma$, and that there exists some finite real number $\eta \in  (\alpha,\beta)$ such that,

\beq
\label{converse}
{\tilde{\mathcal{Z}}}_{T}(t) = \mathcal{O}(|t|^{-r}) \quad\quad r>1\,,
\eeq
when $|t| \rightarrow \infty$ in $\gamma \leq {\rm{Re}}\,t \leq \eta$. Formally, this argument is very similar to defining the asymptotic convergence of a saddle-point equation in terms of its support. Using this theorem the only 
nonvanishing contributions to (\ref{LtransT}) come from the residue contribution, and the asymptotic remainder of the extended fundamental strip. Rewriting the product of poles as a partial fraction for even $T$ we have,

\beq
\prod_{k'=0}^{T} \left(\frac{1} {t+V_{k'}} \right) = \sum_{k'=0}^{T} 
(-1)^{k'} \left(\frac{t + d_{k'}
(\{V_{1}, V_{2}, ... V_{T}\})} {t+V_{k'}} \right)\!\!\,.
\eeq
Integrating (\ref{LtransT}) via parts we have,

\beqa
& & \sum_{k'=0}^{T} (-1)^{k'} \left( \int_{0}^{\infty}  \left( 
\frac{d_{k'}}{t+V_{k'}} \right) e^{zt} dt \right) \\
& = & \sum_{k'=0}^{T}  (-1)^{k'} d_{k'} \, e^{-zV_{k'}}  + 
\mathcal{O}(e^{-z\gamma})\, ,
\eeqa

\beqa
& & \sum_{k'=0}^{T} (-1)^{k'} \left( \int_{0}^{\infty} 
\left(\frac{t}{t+V_{k'}}\right) e^{zt} dt \right)\\ &  = &
\sum_{k'=0}^{T} \frac{(-1)^{k'}}{z} \left( \left[ 
\frac{e^{zt}}{1+\frac{V_{k'}}{t}} \right]^{\infty}_{0} \!\!\!\! +
V_{k'} \int_{0}^{\infty} \!\!\! \frac{1}{(t+V_{k'})^{2}} \, e^{zt} \, dt 
\right) \nonumber \,,\\ \\
& = & \sum_{k'=0}^{T}  (-1)^{k'} V_{k'} e^{-zV_{k'}}  + 
\mathcal{O}(e^{-z\gamma}),
\eeqa

\beq
\label{plaq}
\mathcal{Z}_{T}(z) = \sum_{k'=0}^{T}  (-1)^{k'} \left(d_{k'} + V_{k'} 
\right) e^{-zV_{k'}} + \mathcal{O}(e^{-z\gamma})\,.
\eeq
This relation in (23) gives the general polynomial form of the dual partition function in (\ref{LtransT}), such that it can be meromorphically continued to the left, to large values of $z$ in (\ref{fs1}). What we can conclude from (5) is that if each local term in the physical lattice is of this form of (23) in $z$ then the physical partition function will have a domain of analyticity that asymptotically covers $t\rightarrow+\infty$. It will therefore be implicitly UV regularized according to our Wick-rotated definitions. Similarly, we can write down an asymptotically covered IR regular form for the plaquettes in ${\mathcal{Z}}_{N}(t)$ by meromorphically continuing the converse Mellin transform to the opposite side of the fundamental strip: to the lower bound of the dual. What we have assumed in this is that it is possible to write down the partition function definitions in (5) and (10) as local effective expansions in the gauge coupling on the finite lattice system. This is a similar, but subtly weaker, assumption than existing nonperturbative renormalization approaches in which it is assumed that a local effective expansion can be made in the lattice spacing. 

Since we have implicitly defined the lattice and its dual such that all branch cuts are well-defined in (\ref{Hilberta}) and (\ref{Hilbertb}) the main source of under-determined singularities lies in points that are outside the fundamental strip. Crucially, therefore, we cannot use this converse Mellin mapping procedure to express an arbitrary finite polynomial in $z$ defined over $N$ in a form that is simultaneously free from uncontrolled divergences in both the IR and UV, as defined in our Wick rotated picture. The principle behind the converse Mellin transform theorem allows us to rescale problems by moving the fundamental strip, but it does not allow us to enlarge the fundamental strip. This is the general result of Symanzik as well: an effective expansion in the lattice spacing can only defined with a finite lattice cutoff scale dependence. A nonperturbative renormalization prescription 
like (2) can map lattice data to the limit of vanishing lattice spacing and the continuum, but it does so at the expense of the IR physics of the lattice system. Practically, this has the effect of pushing numerical lattice simulations towards the continuum limit, in order to cover more of the perturbative and nonperturbative scales, and this tends to make simulations computationally more expensive. By defining an analytic procedure that enables us to extend the fundamental strip, not simply move it, we aim to now reduce this cost at a generic level and extend 
the fundamental strip of standard lattice gauge operators analytically. 

\section{Radius of Convergence - comparison with Zeta-function 
renormalization}

Before defining the renormalization group equations for this new picture we draw a comparison between the new polynomial approach we are proposing and $\zeta$-function renormalization, which is similarly constructed from the properties of Mellin transforms. The aim for both this approach and the new approach is to identify the effective action in the form of a product, for the integrand of the path integral that represents the partition function, giving ${\rm{ln}} \mathcal{Z}$ in the form of a sum. As we can see from (\ref{partitionz}) this is complicated in general because the two separate expansions of the integrand in the variables $z$ and $t$ are not defined over
mutually orthogonal bases, and the problem is a noncommuting one. This is the reason in the previous section we have 
only been able to relate the gauge coupling $z$ to the cutoff $t$ through a fairly involved asymptotically correct procedure, and this is the general picture of asymptotic expansion about a renormalization group fixed point. 

In \cite{Zeta}\cite{renorm} the $\zeta$-function, $\zeta$, is defined by a series in the eigenvalues $\{\lambda\}$ of a generalised loop operator $L$, with $z$ a renormalization parameter. The partition function is given by,
\beq
\label{zeta0}
\mathcal{Z}[\phi] = \int \!D\phi \, e^{\int dV \phi L \phi}, \quad
{\rm{ln}} \, \mathcal{Z} = {\rm{ln \, det}} \, Lz\,, 
\eeq
and the $\zeta$-function itself is defined through, 
\beqa
\label{zeta2}
\zeta(t|L) 
& = & \frac{1}{\Gamma(t)} \int_0^\infty x^{(t-1)} e^{-Lx} \, dx\\
\label{zeta}
& = & \sum_{n=0}^{\infty} \lambda_n^{-t} \,,
\eeqa
where $t$ is the regularization parameter, $x=e^{-z}$, and $\phi$ some arbitrary bosonic field. The integrand of the path integral in (\ref{zeta0}) is expressed in the form of the infinite-dimensional characteristic polynomial in the renormalization parameter $z$ of the operator $L$. The $\zeta$-function renormalization method is defined by 
differentiating (\ref{zeta2}) to give a sum for ${\rm{ln}} \mathcal{Z}$. This becomes a well-defined renormalization procedure through Taylor expansion about the point $t=0$,
\beqa
\zeta'(t|L) & = & -\sum_{n=0}^{\infty}\lambda_n^{-t}\,\ln\lambda_n \:\,,\\
\label{Taylor}
\ln\det Lz & \equiv & -\zeta'(0|L)+\zeta(0|L)\ln z\:\,.
\eeqa
The key element in this approach is that (\ref{Taylor}) is simply a definition, for what is otherwise an ill-defined mathematical expression in (\ref{zeta0}) without some form of regularization. From (\ref{Taylor}) we can thus picture that the support of the series of the partition function integrand $Lz$ is UV regularized as a Taylor expansion about some finite value of the cutoff, which in this case happens to be at $t=0$.

The comparison we would like to draw for pedagogy is between the behaviour of the support in this approach and our new approach. The converse Mellin transform theorem of meromorphic continuation that we have used in the previous section in (16) is really a definition of how the support of the transform evolves toward the continuum limit. In this sense, the defining condition in (\ref{Taylor}) is equivalent to the last step in the converse Mellin mapping conditions. Formally, the converse Mellin mapping tells us that the difference between the neighbourhood of the support and the extended region around this region must be monotonically decreasing when this neighbourhood of the support is extended to cover the entire space, and the space is expanded to infinite size. For the Taylor series of the $\zeta$-function definition to be convergent, and for the meromorphic continuation to $t=\infty$ exist, implicitly this difference between the support of the series and the extended region must be finite. A Taylor series defined renormalization procedure is implicitly convergent, upto the cutoff prescription. This is not just true for $\zeta$-function renormalization but this is the same fundamental picture that is developed for the MS scheme and all renormalization schemes: a finite radius of convergence in the cutoff $t$ must exist for any local renormalized expansion \cite{ms}\cite{msbar} as in (28).

The formulation of our dual partition function in (\ref{partitiont}) can be implicitly defined as a commutative product in the variable $x=e^{-z}$ as is the case in the $\zeta$-function scheme (\ref{zeta}). We can therefore similarly factor the dual partition function as a finite product over the eigenvalues of its integrand $\{ \lambda_{k}\}$. The same commutative expansion also exists for the physical lattice in the eigenvalues of its 
integrand $\{ \lambda'_{k}\}$. The underlying purpose of the generalised Wilson loop definitions we have given is that no singularities exist in these definitions as a consequence of the gauge, and we can thus define a local action without any unresolved branch points associated with either the center of the local gauge fields or the noncompactness of the abelian phase. From (5) we can therefore use this feature to split the physical partition function integrand up into a series of contours, and from (10) we can also do the same for the dual partition function. At some critical scale it becomes meaningful relate these two commutative expansions that together define a noncommutative problem. Geometrically this defines a covering between the gauge coupling $z$ and the lattice cutoff $t$. Thus, the physical partition function can be written as an exact finite polynomial expansion in the dual partition function, which defines the finite polynomial analogue of (\ref{Taylor}),

\beqa
& & \!\!\!\!\!\!\!\! \mathcal{Z}_{N}(t) = \int {\rm{d}}g \,
\prod_{k'=0}^{T}{\rm{exp}}\!\left[-\int_{k'\epsilon'}^{(k'+1)\epsilon'} 
L(\bm{n}_{s},z)\,z \,\,ds\,\right] \\
& = & \int {\rm{d}}g  \, \prod_{k'=0}^{T}  e^{\epsilon'} \lambda_{k'} \\
& = & \int {\rm{d}}g  \, \prod_{k=0}^{N} e^{\epsilon} \left(\prod_{k'=0}^{T} 
  e^{\epsilon'/N} \lambda''_{k', k}\right) \\
& = &
\int {\rm{d}}g \,\prod_{k=0}^{N}\,\prod_{k'=0}^{T}\,\,\,
{\rm{exp}}  \bigg[ \int_{k\epsilon/T}^{(k+1)\epsilon/T} \!\!  \int_{k'\epsilon/N}^{(k'+1)\epsilon/N} \nonumber \\ 
& & \quad\quad\quad\quad\quad\quad\quad  L(\bm{n}_{s},\bm{n}_{s'},z,t) \,\,ds'\,ds\,\bigg]\,\, \,,\\
& = & \int {\rm{d}}g \,
\prod_{k'=0}^{T}\prod_{k=0}^{N} \lambda_{k'}^{1/N} \nonumber \,\,
{\rm{exp}}\!\left[\int_{k\epsilon/T}^{(k+1)\epsilon/T} 
L(\bm{n}_{s'},\epsilon') \,\epsilon' \,ds'\,\right]\nonumber \\ \\
& = & \int {\rm{d}}g \,
\prod_{k=0}^{N} \lambda_{k} \,\,
{\rm{exp}}\!\left[-\int_{k\epsilon}^{(k+1)\epsilon} L(\bm{n}_{s'},t) \,t 
\,ds'\,\right] \\
& = & \prod_{k=0}^{N} \left( \lambda_{k} + \mathcal{Z}_{T}(\lambda'_{k} 
e^{\epsilon}) \right)\,,
\eeqa
where $\epsilon' = -t/T$ and $\epsilon = -z/N$. We can further simplify this finite product expansion for $\mathcal{Z}_{N}(t)$ to give a form that will me more useful for evaluating the renormalization group equation,

\beqa
\mathcal{Z}_{N} \, (t) & = & \sum_{k=0}^{N} \Lambda_{k} 
\mathcal{Z}_{T}^{k}(\lambda'_{k} e^{-z}) \\
\label{saddle}
& = & \sum_{k=0}^{N} \Lambda_{k} \int_{\sigma - i\infty}^{\sigma + i\infty}
\frac{1}{(\lambda'_{k})^{t}} \, {\tilde{\mathcal{Z}}}_{T}(t) e^{zt} \, dt\,, 
\eeqa
where $-\alpha < \sigma < \beta$, and $\Lambda_{k}$ are the polynomial coefficients where $\lambda_{k}$ are the eigenvalues of the characteristic polynomial. Implicitly in the $\zeta$-function renormalization scheme two series expansions are used to define the renormalized expansion; the eigenvalue expansion in (\ref{zeta}), and the regularized expansion in $t$ in (\ref{Taylor}). In (\ref{saddle}) we have a renormalized expression of a similar form, with a finite polynomial expansion over $N$ and $T$ where the region of renormalization is specified through the fundamental strip $-\alpha < \sigma < \beta$. Nonperturbative input appears in (\ref{saddle}) in 
the form of the eigenvalues of the physical and dual systems, which can also be used to define the poles $\{ V_{k'} \}$ and the polynomial expansion coefficients $\{ \Lambda_{k} \}$. Unlike the two series that define the $\zeta$-function renormalization it is always possible in our new polynomial approach to change the relative dimensionality of $T$ and $N$ because the scheme is not defined in the continuum. Thus the support of the finite polynomial in $z$ can always be fully enclosed by the domain of analyticity of the rescaled cutoff $t/N$ by effectively changing the covering that relates the two scales. In standard nonperturbative renormalization schemes combinations of gauge fixing and improved operator definition are applied to this effect to improve the analyticity of the effective expansion in the lattice spacing, making this connection between scales. The difference in our new approach is that we are going further, beyond simply improving the gauge elements, to improve the analyticity of the lattice $\beta$-function in the gauge coupling as well, by absorbing nonanalyticities into the generalised Wilson loop operator definitions associated with the transition between different topological sectors at $\theta=\pi$.

\section{Renormalization Group Equation}
By making a comparison between $\zeta$-function renormalization and a noncommutative Lattice Gauge Theory definition we have argued in Section V that we can define a physical lattice partition function, in a renormalized form, as a finite order polynomial in its dual, (35). We have also argued separately in Section IV that in order for the Laplace transform of the dual partition function to be meromorphically continuable to $t\rightarrow\infty$ that the dual must have a quite specific polynomial form as a function of the gauge coupling $z$ (23). What we would like to do now is to combine these two results to further define a nonperturbative renormalization prescription that has been implicitly UV regularized. To give a usual renormalization group equation like (1), we will now write down an effective action of the same form as (\ref{Taylor}), but instead defined as a finite order characteristic polynomial. This will be found by differentiating the logarithm of (37), and then UV regularizing this relation by substituting (23) into (37).

As we have shown in Section IV for each term in (5) we only know the analytic continuation properties of $z$ up to the bounds on the integration over $t$. The polynomial expansion will have cusps at the boundaries of this region, as we argued in Section V to define (35). In order to define the renormalization group equation we must, therefore, sum over the branch cuts about these singular points at the cusps. This can be simply achieved by differentiation with respect to the integrand eigenvalues of ${\mathcal{Z}}_{T}(z)$, \cite{cusp}. The renormalization group equation for the effective action is then given, in the same form as (\ref{Gell}), by the differentiation of this local effective action with respect to the local scale transformation parameter $z$, 

\beqa
& & \frac{\partial}{\partial z } \left( \sum_{k'=0}^{N}
\frac{\partial }{N\partial \lambda'_{k'}} \ln \mathcal{Z}_{N} 
\!\!\left(\mathcal{Z}_{T}(\lambda'_{k}e^{-z})\right) \right) \\ & = &
-\frac{1}{\mathcal{Z}^{2}_{N}(t)} \frac{\partial 
\mathcal{Z}_{N}(t)}{\partial z }
\sum_{k=0}^{N-1} k\Lambda_{k+1} 
\mathcal{Z}^{k}_{T}(\lambda'_{k}e^{-z})\frac{\partial 
\mathcal{Z}_{T}(\lambda'_{k} e^{-z})}{\partial \lambda'_{k} }
\nonumber \\
& + & \!\!\! \sum_{k=0}^{N-2} k(k+1)\Lambda_{k+2} 
\mathcal{Z}^{k}_{T}(\lambda'_{k}e^{-z})
\frac{\partial \mathcal{Z}_{T}(\lambda'_{k} e^{-z})}{\partial \lambda'_{k} }
\frac{\partial \mathcal{Z}_{T}(\lambda'_{k} e^{-z})}{\partial z }\nonumber
\\ & \times & \!\!\! \frac{1}{\mathcal{Z}_{N}(t)} + \frac{1}{\mathcal{Z}_{N}(t)}
\sum_{k=0}^{N-1} k\Lambda_{k+1} \mathcal{Z}^{k}_{T}(\lambda'_{k}e^{-z}) 
\frac{\partial^{2}
\mathcal{Z}_{T}(\lambda'_{k} e^{-z})} {\partial z \partial \lambda'_{k} 
}\nonumber \,.\\
\eeqa

Substituting in (\ref{plaq}), the UV regularized solutions of the renormalization group equation thus appear as the solutions of,

\beq
\frac{A -  \mathcal{Z}_{N}(t)[\,B\,+\,C\,]}{\mathcal{Z}_{N}^{2}(t)}= 0 \,,
\eeq
with,

\beqa
A & = & \left( \sum_{j=0}^{N-1} j\Lambda_{j+1} 
\mathcal{Z}^{j}_{T}(\lambda'_{k}e^{-z}) \right)
\left( \sum_{k=0}^{N-1} k\Lambda_{k+1}  \mathcal{Z}_{T}^{k} 
(\lambda'_{k}e^{-z}) \right) \nonumber \\
& \times & \sum_{j'=0}^{T} (-1)^{j'}\, (\lambda'_{j}e^{-z})^{V_{j'}}
\left[ (d_{j'}+ V_{j'})V_{j'} \right] \nonumber \\ & \times & 
\sum_{k'=0}^{T} (-1)^{k'} \,
(\lambda'_{k} e^{-z})^{V_{k'}} \left[ (d_{k'}+ V_{k'}) 
\frac{V_{k'}}{\lambda'_{k}}  \right]\,,
\eeqa

\beqa
B & = &
\sum_{k=0}^{N-2} k(k+1)\Lambda_{k+2} \mathcal{Z}^{k}_{T}(\lambda'_{k}e^{-z}) 
\nonumber \\
& \times & \left( \sum_{k'=0}^{T} (-1)^{k'} \, (\lambda'_{k}e^{-z})^{V_{k'}} 
\left[ (d_{j'}+V_{j'}) V_{j'} \right]\right)\nonumber \\
& \times & \left( \sum_{k'=0}^{T} (-1)^{k'} \, (\lambda'_{k} 
e^{-z})^{V_{k'}}
\left[ (d_{k'}+V_{k'})\frac{V_{k'}}{\lambda'_{k}} \right] \right)\!\!,
\eeqa

\beqa
C & =& \sum_{k=0}^{N-1} k\Lambda_{k+1} 
\mathcal{Z}^{k}_{T}(\lambda'_{k}e^{-z})\nonumber \\
& \times & \left( \sum_{k'=0}^{T} (-1)^{k'} \, (\lambda'_{k}e^{-z})^{V_{k'}}
\left[ (d_{k'}+ V_{k'})\frac{V_{k'}^{2}}{\lambda'_{k}} \right] \right)\!\!,
\eeqa

Since the noncommuting Lattice Gauge Theory formalism is holomorphic in $z$ we can equate the terms of (40) to finite order in $z$ to identify the solutions of (40). The solutions of the nonperturbative renormalization group equation therefore appear as either the zeroes of the polynomial in (37), or as the solution of a set of coupled equations in the expansion coefficients, 

\beqa
\label{1}
& & \Lambda_{j}^{2} ( 1 - \Lambda_{j-1}) - \Lambda_{j+1}\Lambda_{j-1} = 0\,.
\eeqa

The zeroes of the polynomial in (37) are quite difficult to solve for directly, from nonperturbative input, since standard numerical lattice analyses will yield the physical lattice eigenvalues, but not the poles of dual. However, the coupled equations in (\ref{1}) are much simpler to resolve. These can be resolved implicitly by the relative choices of $T$ and $N$ we specify for the new operator definition in (8), as we will now show. 

These relations in (44) define a saddle-point equation about each local term on the physical lattice from nonperturbative input, and this gives the radius of convergence of the renormalization  throughout the lattice. 
This follows from the arguments for the support we developed in Section V. In essence, therefore, these coupled equations define the analogue of the arbitrary physical reference scale in (\ref{Gell}) $\mu$, about which the effective generator expansion is defined. They define nonperturbatively the external momenta of the states the local 
gauge operators interpolate between, implicitly, in terms of $t$. The central difference from (2) in this picture is that these external states, and cutoff prescription, are defined locally for each site. A gauge fixing prescription cannot be given for the entire lattice system in this picture because we have included the freedom to sum over topological sectors in the operator definitions. The operators themselves are not necessarily analytically connected between the local choices of branch coming from the generalised Wilson loop operators in Section IV, without further local information. This is not an inherent problem, and the continuum theory will correspond to a single topological sector, but the lattice $\beta$-function is modified in this picture.

To solve (44) we firstly need to identify the operator $V(\bm{n}_s)$ in (6) from the standard Lattice Gauge Theory operators of a given numerical ensemble, using,

\beqa
\label{gauge}
V(\bm{n}_s) & = & U(s;s+1) - B_g(\bm{n}_s)\,z \\
	    & = & U(s;s+1) - S(s;s+1) U(s;s+1) \nonumber \\
& \times & S(s;s+1)^{-1}\,z \,,
\eeqa
where $\{S(s;s+1)\}$ are the similarity transforms that diagonalise each local site separately within each configuration of a numerical ensemble. What we next need to do is to evaluate the $T$-gauge dependence of projecting these operator elements onto the dual in (9). To evaluate this projection $V(\bm{n}_{s'})$, using standard techniques, we can project the usual lattice ensemble onto itself, following from the definitions in (\ref{Hilbertb}),

\beqa
& & \langle \bm{n} | V( \bm{n}_{s'} ) | \bm{n}_{s} \rangle \nonumber \\ & = &
\sum_{(s,s')}^{N \otimes T} \,\, \sum_{\sigma \in G}
\lambda''_{ss'\sigma}(\bm{n})\frac
{\langle \bm{n}\oplus \bm{1}_{s\sigma} \oplus \bm{1}_{\sigma s'}| \bm{n}_{s} 
\rangle}
{\langle \bm{n}_{s}| \bm{n}_{s'} \rangle} \langle \bm{n}| \bm{n}_{s'} 
\rangle \,. \nonumber \,,\\
\eeqa
We can then calculate the product of two of these matrix elements to completely remove the $T$-gauge dependence,

\beqa
& & \!\!\!\!\!\!\!\! \langle \bm{n} | V( \bm{n}_{s'} ) | \bm{n}_{s} \rangle \, \langle \bm{n}_{s} 
  | V( \bm{n}_{s'} ) | \bm{n} \rangle =
\langle \bm{n} | V^{2} ( \bm{n}_{s'} ) | \bm{n} \rangle \,\\
& = & \sum_{(s,s')}^{N \otimes T} \,\, \sum_{\sigma \in G}
\lambda''_{ss'\sigma}(\bm{n})\frac
{\langle \bm{n}\oplus \bm{1}_{s\sigma} \oplus \bm{1}_{\sigma s'}| \bm{n}_{s} 
\rangle}
{\langle \bm{n}_{s}| \bm{n}_{s'} \rangle} \langle \bm{n}| \bm{n}_{s'} 
\rangle \, \nonumber \\
& \times & \sum_{(s,s')}^{N \otimes T} \,\, \sum_{\sigma \in G}
\lambda''_{ss'\sigma}(\bm{n})\frac
{\langle \bm{n}_{s} \oplus \bm{1}_{s\sigma} \oplus \bm{1}_{\sigma s'}| 
\bm{n} \rangle}
{\langle \bm{n} | \bm{n}_{s'} \rangle} \langle \bm{n}_{s}| \bm{n}_{s'} 
\rangle \, \nonumber \\ \\
& = &
\label{Tgauge}
\sum_{(s,s')}^{N \otimes T} \,\, \sum_{\sigma \in G}
\lambda''_{ss'\sigma}(\bm{n})
\langle \bm{n}\oplus \bm{1}_{s\sigma} \oplus \bm{1}_{\sigma s'}| \bm{n}_{s} 
\rangle  \nonumber \\ \,
& \times & \sum_{(s,s')}^{N \otimes T} \,\, \sum_{\sigma \in G}
\lambda''_{ss'\sigma}(\bm{n}) \langle \bm{1}_{\sigma s'}| \bm{n} \rangle \,.
\eeqa

This then gives a gauge-fixed lattice ensemble defined in the eigenvalues of the projection onto $T$. This is a projection of the noncommutative Lattice Gauge Theory operators onto a single topological sector. To be clear, what we are defining practically is a nonlocal gauge prescription in (46), for which we then define an inner product over $T$ on each of the matrix elements of these new gauge operators in (50) in order to fix this gauge. For these new elements to be topologically complete we simply need to reinsert the discrete center of the projection, by multiplying each of our new operator elements by the ${\rm{dim}}\,G$-fold roots of unity. The inverse Laplace transform of our new procedure is therefore given by,

\beqa
\widetilde{\mathcal{Z}}_{N}(z)
& = & \prod_{s=0}^{N} \prod_{s''=1}^{{\rm{dim}}\,G} \,\, 
\prod_{s'''=1}^{{\rm{dim}}\,G}
\frac{1}{\root {\rm{dim}}\,G \of {(\,z + V_{s} \,)}} \,\, 
\delta_{s''}^{s'''} \\
& = & \prod_{s=0}^{N} \prod_{s''=1}^{{\rm{dim}}\,G} \frac{1}{ z^{2} + 
\alpha_{s''}(V^{2}_{s}) } \,.
\eeqa
where $\{ \alpha\}$ are the eigenvalues that completely define algebra of the gauge field elements in (3). We still have to take the branch of a square root in order to relate the elements in (5) to the partition function in 
(\ref{partitionz}) but will now know the sign of the original matrix elements in (\ref{Tgauge}). If the sign of the 
matrix element in (\ref{Tgauge}) for a given physical site $s$ is positive then we can use this to define the relative orientations of the branches for the square root. If the sign of the eigenvalue matrix element in (\ref{Tgauge}) for a different physical site $s$ is then negative, then we simply multiply the gauge factor elements in (5) for this particular site by the square root of unity to maintain the consistency of the branch throughout the lattice system. 

With this choice of gauge fixing, the partition function in (35) is necessarily quadratic in $z$ and there are no unresolved branch points on the finite lattice system in making this choice. Practically, this means that (44) is always satisfied. The zeroes of the polynomial in (37) can therefore be used to completely determine the nature of the singularities in the lattice spacing, and in particular the criticality of the renormalization group fixed points as the lattice system evolves into a continuum theory, following from the arguments for the support in Section V. What we have done here is to define how to remove singularities in the lattice spacing from an expansion in the gauge coupling, such that the singularities in the new modified gauge coupling have a prescribed convergence to the continuum limit, following from the converse Mellin mapping theorem.

\section{Applications}

From a noncommuting Lattice Gauge Theory formalism we have defined a UV regularized renormalization procedure, in Section V, based on a generalisation of the $\zeta$-function renormalization scheme. Then, introducing a nonlocal gauge-fixing prescription, in Section VI, we have further identified that the solution of the renormalization group equation for this approach comes from the zeroes of a characteristic polynomial equation. In general, the physical problems for which we would like to apply nonperturbative lattice renormalization techniques come with a variety of nonlocal taste symmetries for the partition function, and also different taste and spin symmetries for the operator elements of physical interest. The procedure we have defined in (52), from the arguments in Sections V and VI, is at a generic level for Wilson loop operators of arbitrary taste. What we would now like to do is treat some specific operators of physical interest with this procedure, and, in particular, identify whether the characteristic polynomial zeroes can be uniquely categorised for these different physical problems. To do this, we first make a pedagogical comparison with the RIMOM nonperturbative renormalization prescription.

The renormalization parameter for an arbitrary operator (defined as a function of the Dirac matrix $\gamma_{j}$) is defined in the RIMOM scheme for (2) through, 
\beq
\frac{Z_{O}( \mu a, g(a) )}{ Z_{\psi}( \mu a, g(a) )} \, \Gamma_{O} (p a)\, |_{p^{2}=\mu^{2}} \equiv 1\,,
\eeq
where $Z_{\psi}$ is the renormalization constant for the interpolating quarks, $\Gamma_{O} (p a)$ is the forward amputated Green's function for this operator computed between quark states of momenta $p$, which itself is derived from taking the trace over the colour and spin states of the nonamputated Green's function $G_{O}(pa)$ and quark propagators $V(pa)$ found nonperturbatively found from inverting the lattice Dirac operators,  
\beq
\Gamma_{O} (p a) = \frac{1}{12} {\rm{Tr}} \Big( \,V(pa)^{-1} G_{O}(pa) \, V(pa)^{-1} \, P_{O}(a) \, \Big)\,, 
\eeq
where $P_{O}(a)$ is a projection operator for $\gamma_{j}$ on the lattice. The nonamputated Green's function $G_{O}(pa)$ in this prescription is found in the usual way on the lattice by summing over the total number of configurations $\#$ of the lattice Monte Carlo ensemble. Where the momentum space dependence is found through an explicit Fourier transform of the interpolating operators,
\beqa
G_{O}( x, y ) & = & \langle \psi(x) O_{\gamma_{j}}(0)\bar{\psi}(y) \rangle \\ 
& = & \frac{1}{\#}\sum^{\#}_{n=1} V_n(x \vert 0) \gamma_{j} V_n(0 \vert y) \,,\\
G_{O}(pa) & \equiv  & \int d^4 x \, d^4 y  e^{-i \,( p \cdot x- p \cdot y)}
G_O( x ,  y ) \\ 
& =&\frac{1}{\#} \sum_{n=1}^{\#} V_n(p \vert 0)
 \gamma_{j}  \Bigl( \gamma_5 V^\dagger_n(p \vert  0) \gamma_5
 \Bigr) \,.
\eeqa
We have introduced a suggestive notation here for the quark propagators $\{V\}$ that relates directly to the noncommutative operator definition in (7). It is in fact, possible for us to view the noncommutative operator elements as a more generalised version of the quark propagators. Since $T$ is defined arbitrarily the dual definition can be chosen so that the noncommutative elements $V$ are identically the quark propagators. The extended basis dimension $T$ would then span all the lattice sites, with all-to-all nonlocal correlations. 

In fact, using the definition in (58) the defining relation for RIMOM in (54) is of the same general form as the $T$-gauge fixing operator definition in (50). In (58) a local inner product is defined for the quark propagators over the taste and spin indices, by averaging over the Monte Carlo ensemble, which is then projected nonlocally in (54). What we have in (50) is that a local inner product is defined for the noncommutative elements, which is then generalised nonlocally through an analytic definition of the relation between the local bases, to give a new set of operators. The renormalization definition in (53) is nonlocal in the sense that the matrix elements that form 
the nonamputated Greens function have been obtained from averaging, and represent the expectation of the support of the Greens function. Our method essentially defines this procedure analytically without explicitly multiplying out the $N\! \times \!N$ quark propagators. We consequently have $N$ such relations of the form of (50) each coming from the nonlocal gauge-fixed noncommutative operators for each physical lattice site, all of which can be used to infer information about the convergence to the continuum limit of the renormalization prescription. 

\subsection{Ward Identities}
A key question is whether or not the new operator formalism satisfies the chiral Ward identities for the vector and axial vector currents. The basic answer is yes, because (53) satisfies these identities, which is shown explicitly in \cite{wi} and \cite{rimom}. Provided the second term vanishes in the following relation, which is only true at large momenta for the axial vector current, then both of the Ward identities are simply satisfied by tracing over the matrix products in (53),

\beqa
Z_{V^L}\Bigl( \Lambda_{V^L_\rho}(p) + \!\!\!\!\!\!\!\!\!\! & & q_\mu \frac{\partial}{\partial q_\rho}
\Lambda_{V^L_\mu}(p+\frac{q}{2},p-\frac{q}{2})\vert_{q=0}\Bigr) \nonumber \\
& = & -i \frac{\partial}{\partial p_\rho}V^{-1}(p)\,, 
\eeqa
where,
\beq
\Lambda_{O}(pa) = V(pa)^{-1} G_{O}(pa) \, V(pa)^{-1} \,.
\eeq

What we are doing that is very different, is defining this relation locally through the operator prescriptions, so 
that it applies to each plaquette separately. At least, the support of each plaquette is defined separately but all the local Ward Identity relations are analytically connected, through the operator prescriptions we have given in Section VI. So, even though we are not evaluating this relation in (59) in the usual functional sense, by averaging the lattice operators over all sites and over all configurations in the statistical ensemble, the local operators still represent a disconnected expansion using all this information from all the lattice sites.

To replicate the momentum transfer in the second term in our formalism, following (57), we can separate the two operator elements involved in the inner product in (48) by an arbitrary number of physical lattice sites to form an $N'\!\times \! N'$ basis element on the physical lattice \cite{N*}. From (59) what we are interested in is to satisfy the continuum Ward Identity. This will constitute evaluating the first derivative of the second term, in the limit that the momentum transfer is zero. Therefore, from the meromorphic properties defined in Section V, what we will be interested for the new formalism is the analyticity of the holomorphic variable $t$ as we move analytically between spatial sites in the $N'\!\times \! N'$ element and also the limit $N'\rightarrow 1$ when the momentum transfer is consequently zero. The equivalent of the quark propagators in this basis are given by,

\beqa
& & \langle \bm{n} | V( \bm{n}_{s'} ) | \bm{n}_{S} \rangle \nonumber \\ & = &
\sum_{(S,s')}^{N \otimes T} \,\, \sum_{\sigma \in G}
\lambda''_{Ss'\sigma}(\bm{n})\frac
{\langle \bm{n}\oplus \bm{1}_{S\sigma} \oplus \bm{1}_{\sigma s'}| \bm{n}_{s} 
\rangle}
{\langle \bm{n}_{S}| \bm{n}_{s'} \rangle} \langle \bm{n}| \bm{n}_{s'} 
\rangle \,. \nonumber \\
\eeqa

where $S=s\otimes s\otimes s ... s$ with this being the direct product over $N'$ physical sites. We have already discussed the analyticity of $z$ when a $Z(2)$ symmetry is realised on the operator elements in (50). In this case associated with the branch cut in $z$ each local sites picks up a phase factor from the nonperturbative choice of branch. In practice, this means that the vortices defined for the $N\otimes T$ elementary plaquettes have an associated sign: either an clockwise or anticlockwise sense on $N\otimes T$. In general what will happen is that when we differentiate the $N'\!\times \! N'$ element with respect to $t$ it will pick up this nonperturbative phase factor. Branches that traverse this operator element in the opposite spatial direction will have this phase factor inverted. We can insert a Dirac matrix projection operator into the inner product for the gauge fields 
in (48) to project onto a colour/spin sector. If this choice of Dirac matrix is not affected by Wick rotation then this will have no effect on the choice of branch for the path. However, if we insert $\gamma_{5}$ the reverse spatial path is specified over different matrix elements in (48). Consequently the vanishing of the second term in (59) is an identity in our formalism because of the cancellation of paths for the vector Ward Identity, but for the axial Ward Identity the forward and reverse paths are never equivalent which gives rise to a finite phase state in the $N\otimes T$ vortices and massless mode.

\subsection{Divergent composite operators}
The converse Mellin mapping theorem, we discussed in Section IV, is very important for the new features of the analysis of our prescription. It tells us how we can meromorphically continue the gauge elements in the lattice cutoff $t$. This tells us, essentially, how to define the lattice gauge plaquettes semianalytically in the gauge 
coupling $z$ such that they will remain UV regular under $z\rightarrow z'$. What the theorem also tells us is how the lattice evolves as the physical size, and number of poles on the physical lattice, is increased, from (23). This convergence prescription applies whether we consider expanding the volume of a small region of physical sites within an existing lattice system, or the evolution of one lattice volume to another closer to the continuum. The theorem gives that the difference between the support of (23) and the neighbourhood of the support must be monotonically decreasing as the size is increased, and this is the basis for our renormalization approach in Section VI. We can average over the local nonperturbative supports by diagonalising the lattice as in (36). This is effectively done in (58) by using the inverse of the lattice Dirac matrix as an interpolating quark propagator. As with the support of the eigenvalues of the Dirac matrix the zeros of the characteristic polynomial in (36) should converge monotonically to the real axis of the complex plane. Only in this case the expansion parameter is the partition function of the dual which gives the local normalisation of the physical plaquette in the lattice cutoff $t$.

We can apply this property to look at the problem of operator mixing induced by explicit chiral symmetry breaking on the lattice. This can lead to two sorts of divergences as the continuum limit is approached for four-fermion operator terms \cite{4f1}-\cite{4f3}; logarithmic divergences for $\Delta S=2$ processes, such as those associated with the $K^{0}-{\bar{K}}^{0}$ mixing amplitude \cite{4f4}\cite{4f5}, and power divergences from the mixing with lower dimensional operators for the penguin operators appearing in the $\Delta S=1/2$ Hamiltonian \cite{4f6}.
The four-fermion operators for these cases are, 
\beqa 
O^{\Delta S=2} & = & \Bigl(\bar s \gamma^\mu (1-\gamma_5) d \Bigr) \Bigl(\bar s \gamma_\mu (1-\gamma_5) d \Bigr) \,,\\
O^{ \Delta S=1/2} & = & ( \bar s \gamma_\mu ( 1 - \gamma_5 )d ) \sum_i (\bar q_i \gamma^\mu ( 1 + \gamma_5 ) q_i) \,.
\eeqa
where the lattice renormalization prescription for these operators and leading mixed terms follows, 

\beq 
O^{\Delta S=2}_{{\rm cont}} = Z_O \Bigl( O^{\Delta S=2}_{{\rm
latt}} + Z_{1}(\overline{s}d)(\overline{s}d) + Z_{2}(\overline{s}\gamma_{5}d)(\overline{s}\gamma_{5}d) ... \Bigr) \,,
\eeq
\beq
O^{\Delta S=1/2}_{{\rm cont}} = Z_O O^{\Delta S=1/2}_{{\rm latt}}+ Z_1 (\bar s d )+ Z_2 (\bar s  \gamma_5 d )+ \dots \,\,. 
\eeq
In this case the noncommutative matrix elements will be defined over a three flavour, three colour component group 
manifold, such that $G \simeq SU(N_{f}) \otimes SU(N_{c})$. To solve for the three renormalization constants one can, in the RIMOM scheme, separately evaluate the subspace projection of these different operators for the quark interpolating operator definitions in (55)-(58). Basically what is different in the new approach is that the local renormalization factors, found from constructing and solving the characteristic polynomial of these noncommutative operators in (50), will implicitly converge to these three values. There is no need to separately evaluate the subspace projections. Each of the convergent solutions, as a function of increasing system size, will be evident from the convergence of the zeros in the complex plane towards the real axis. The mixed term contributions can be identified from the symmetries of the zeros, and the different convergence rates, as a function of $N$, can be used to infer the relative dominance of the contributions in the continuum limit.

\section{Summary}

We have considered how the continuum limit can be approached for a finite lattice system, by rewriting standard Wilson operators in a noncommutative gauge operator formalism so that they are diagonal in the gauge coupling. We have considered in particular how the lattice spacing evolves in this basis. From the holomorphic and meromorphic properties of our new definitions we have identified the range in the nonperturbative lattice cutoff, over which, a finite polynomial expansion in the gauge coupling can be exactly meromorphically continued to the UV. This new approach strongly resembles a finite Hilbert space generalization of the $\zeta$-function renormalization scheme for noncommutative quantum field theories. Thus, in principle, through our choice in a nonlocal gauge-fixing prescription in this noncommutative basis the gauge operators can be continued arbitrarily close to the continuum limit in the gauge coupling, and the gauge elements can be related directly to the continuum gauge coupling expansions. We are essentially ameliorating the singularities of the lattice $\beta$-function in this new choice of basis.  

Practically, this new approach has the advantage that it alleviates the need for the extrapolation of bare lattice variables with an effective field theory treatment, whose symmetries may not necessarily correspond exactly to those realised because of the lattice spacing dependence. This has been a central motivation for the new work. This effect is particularly important for CP-violating processes with four-fermion interaction terms, which have a strong operator mixing component in the renormalization group flow of the lattice operators to the continuum, and a nonlocal, IR content. More importantly, though, the characteristic polynomial zeroes that define the solution of the new lattice renormalization group equation, in this approach, indicate the presence of all and any singularities that develop in the continuum limit of the nonperturbative evaluation realised in practice. Direct information is thus obtained both for the renormalization group fixed point and the anomalies, as we have explicitly discussed for the case of $\Delta S=2$ and $\Delta S=1/2$ processes. These latter artefacts may occur, for example, from the topological anomalies that develop as a consequence of the explicit chiral symmetry breaking of the lattice prescription \cite{renorm}, or from the efficiency of the covering of a statistical ensemble. One cannot otherwise strictly speaking extrapolate an effective theory through these nonanalytic singularities, which yield nonrenormalizable multiplicative corrections, without fully knowing the IR behaviour of the nonperturbative lattice $\beta$-function. This feature tends to lead to a large separation of momentum scales being employed in conventional nonpertubative evaluations, which is computationally expensive, and an aim has been to seek a means to reduce this cost at a generic level. Knowing the location of these singular features, through this new prescription, it can be envisaged to remove these nonlocal effects from subsequent studies through the inclusion of counter terms in the action, implemented through the operator prescription, as is done with existing renormalization group improved lattice actions and treatments.

\end{document}